\documentclass[12pt,letterpaper,dvips]{article}
\usepackage[letterpaper,margin=0.75in]{geometry}
\newcommand{\emphx}[1]{``\emph{#1}''}

\begin{document}

\begin{center}\textbf{\large A Comment on \emphx{A New Degree of Freedom For
      Energy Efficiency of Digital Communication Systems}}\smallskip

  \let\thefootnote\relax\footnotetext{Submitted to \emph{IEEE Transactions on
      Communications} on 9 May 2017}

  Pavel Loskot\let\thefootnote 1\footnote{Swansea University, Swansea SA1 3XL,
    United Kingdom. E-mail: p.loskot@swan.ac.uk, Tel.: +44 1792 602619}
\end{center}

\begin{abstract}
  This comment recalls a previously proposed encoding scheme involving two
  synchronized random number generators (RNGs) to compress the transmission
  message. It is also claimed that the recently proposed random number
  modulation (RNM) scheme suffers considerably from the severe error
  propagation, and that, in general, the overall energy consumption is
  minimized when all information bits are transmitted as fast as possible with
  the minimum latency.
\end{abstract}

The paper [1] proposes a random number modulation (RNM) to trade-off the
transmission latency with the energy efficiency. It uses a RNG at the
transmitter and a RNG at the receiver. These two generators are synchronized to
produce identical sequences of $B$ random bits in every time slot. If the
random $B$ bit sequence at the transmitter in a given time slot matches one of
the $M<2^B$ binary sequences representing the $B$ information bits to be
transmitted, the corresponding $M$-ary modulation symbol is transmitted in that
time slot.

Previously, the use of pseudo-random sequences to exploit the
degrees-of-freedom of the communication link was advocated in the paper [2].
Therein, an encoding scheme using two synchronized RNGs at the transmitter and
at the receiver is devised to greatly reduce the number of message bits that
need to be actually transmitted over a feedback link. The RNGs generate $2^C$
random sequences at each time slot until one of the random sequences matches
the feedback message, and then $C\ll \log_2 N!$ bits are reported to the
transmitter where $N!$ is the number of permutations of $N$ bits. The value of
$C$ in [2] and the value of $B$ in [1] can be set to trade-off the transmission
latency.

More importantly, the scheme in [2] assumes an ideal noise-less feedback
channel which is a common assumption adopted in the literature. The compressed
feedback message can be then perfectly recovered at the transmitter. On the
other hand, the RNM signaling in [1] cannot reliably identify the time slots
containing the transmitted $M$-ary modulation symbols. Consequently, the RNM
scheme in [1] suffers from three types of symbol errors having the comparable
probabilities. The first type of symbol error inserts an incorrect non-zero
modulation symbol into the received sequence even though no symbol was
transmitted in that time slot. The second type of symbol error assumes an empty
time slot, so the transmitted non-zero modulation symbol is incorrectly
deleted. The third type of symbol error interchanges the transmitted non-zero
symbol for another incorrect non-zero symbol. The probability of symbol error
(11) in [1] can be used to account for this third type of symbol errors.
However, calculating the overall symbol error probability involving all these
three types of symbol errors is rather non-trivial. The random insertion and
deletion of modulation symbols in the transmitted sequence corresponding to the
first two types of symbol errors causes a misalignment of the transmitted and
the received symbol sequences which gives rise to yet another type of symbol
errors. This misalignment will produce the prolonged error propagations, and
the overall error performance will be significantly worse than reported in the
paper [1]. The consideration of error propagation is even more important in
multiantenna and multiuser scenarios with co-channel interference.

Hence, the symbol errors in [1] are correlated, and the optimum detector at the
receiver have to consider the complete received sequences which affects the
transmission latency. In order to avoid the detection complexity, a sub-optimum
two-stage detection scheme for signalings employing $M$-ary modulation
constellations extended with a zero symbol was considered in the paper [3]. The
signaling in [3] assumes that there is exactly one non-zero $M$-ary modulation
symbol transmitted within a predefined number of time slots. The non-coherent
detection is used to find the likely transmission time slot which is followed
by a conventional coherent symbol detection. In this case, there is no symbol
error propagation, so the performance analysis is more straightforward.

It is also useful to make other general comments about the signaling schemes
employing a zero modulation symbol. The frequent antenna switching is highly
undesirable, especially when it is done at the level of individual modulation
symbols. Such switching considerably expands the signaling spectrum as shown in
[3] which affects the spectral efficiency. In addition, the electronic circuits
have non-zero turn-on and turn-off times which is limiting the achievable
switching frequency. In terms of the energy efficiency, it is shown in [3]
that, in general, the faster the transmission, i.e., the smaller the latency,
the smaller the overall energy consumption consisting of the overhead energy
and the energy expended for the radio-frequency (RF) transmissions. However, it
is correct that when only the RF portion of the overall energy consumption is
considered, one can trade-off the latency for the energy consumption as
discussed in [1].

\smallskip

\end{document}